\documentclass{article} 

  
\usepackage[margin=1in]{geometry} 

\usepackage{times}  
\usepackage{helvet}  
\usepackage{courier}  

\usepackage[hyphens]{url}  
\usepackage{graphicx} 
\usepackage{caption} 
\usepackage{algorithm}
\usepackage{algorithmic}
\usepackage{bm}
\usepackage{amsfonts}

\usepackage{newfloat}
\usepackage{listings}
\usepackage{amsmath}
\usepackage{amsthm}

\DeclareCaptionStyle{ruled}{labelfont=normalfont,labelsep=colon,strut=off} 
\floatstyle{ruled}
\newfloat{listing}{tb}{lst}{}
\floatname{listing}{Listing}
\title{Scaling up Dynamic Edge Partition Models via Stochastic Gradient MCMC}
\author{Sikun Yang\thanks{School of Computing and Information Technology, Great Bay University, 523000 Dongguan, China; Great Bay Institute for Advanced Study, Great Bay University; Dongguan Key Laboratory for Data Science and Intelligent Medicine, Great Bay University. sikunyang@gbu.edu.cn}
\and Heinz Koeppl\thanks{Technische Universit\"at Darmstadt, Darmstadt, Germany. heinz.koeppl@bcs.tu-darmstadt.de}
}

\date{}

\usepackage{bibentry}

\begin{document}

\maketitle

\begin{abstract}
The edge partition model (EPM) is a generative model for extracting an overlapping community structure from static graph-structured data. In the EPM, the gamma process (GaP) prior is adopted to infer the appropriate number of latent communities, and each vertex is endowed with a gamma distributed \emph{positive} memberships vector. Despite having many attractive properties, inference in the EPM is typically performed using Markov chain Monte Carlo (MCMC) methods that prevent it from being applied to massive network data. In this paper, we generalize the EPM to account for dynamic enviroment by representing each vertex with a positive memberships vector constructed using Dirichlet prior specification, and capturing the time-evolving behaviour of vertices via a Dirichlet Markov chain construction. A simple-to-implement Gibbs sampler is proposed to perform posterior computation using Negative-Binomial augmentation technique. For large network data, we propose a stochastic gradient Markov chain Monte Carlo (SG-MCMC) algorithm for scalable inference in the proposed model.  The experimental results show that the novel methods achieve competitive performance in terms of link prediction, while being much faster.
\end{abstract}

\section{Introduction}
Analysis of graph-structured data has been an active research area because of its many applications in social network analysis~\cite{pnas_a}, recommender system~\cite{Gopalan, Charlin, Cao18} and knowledge graph embedding~\cite{KnowBase, KnowGraph, NDKG}. Over the last decade, many efforts have been dedicated to modelling graph-structured data, such as the stochastic block model (SBM)~\cite{MMSBM}, the latent feature relational model (LFRM)~\cite{LFRM}, the edge partition model (EPM)~\cite{EPM} etc. Among these methods, the EPM has drawn much attention because of its appealing properties: (1) the gamma process prior is adopted to automatically  infer the appropriate number of latent communities; (2) each vertex is represented by a gamma distributed \emph{positive} memberships vector that is highly interpretable, and allows us to extract an overlapping community structure from the given network; (3) the EPM model enjoys full local conjugacy and admits a simple-to-implement MCMC algorithm using data augmentation and marginalization strategy for posterior computation. 
The EPM has also been extended to multi-relational graph learning~\cite{TopicKG}, deep generative graph modelling~\cite{DGMR}, inductive matrix completion~\cite{Piyush17, Zhao}, and dynamic network modelling~\cite{BPTD,DPGM,ICML-18,UAI-20}.  While the EPM and its extensions achieve competitive performance in various applications, inference in these models is done via MCMC methods that notoriously mix slowly and scale poorly to large dataset in practice. Real temporal graph-structured data, such as user-user interactions in social networks or user-item ratings in recommender systems, easily run into millions of vertices and edges. Hence, it is highly desirable to develop a scalable inference algorithm for the EPM and its various extensions. 

In this work, we propose a novel generative model that extends the gamma process edge partition model (GaP-EPM) to model temporal assortative graphs by endowing each vertex with a positive memberships vector, constructed using Dirichlet prior specification, which retains the expressiveness and interpretability of the original EPM. Specifically, the new model utilizes a Dirichlet Markov chain to capture the smooth evolution of the vertices' memberships over time. 
Unlike the original EPM construction, the ideal number of latent communities is adaptively inferred using the hierarchical beta-gamma prior specification~\cite{NBP}. Note that the edge partition model has been extended to model temporal networks via evolving vertices memberships through gamma Markov chain construction~\cite{DPGM}. However, the gamma priors construction for the original GaP-EPM can disturb the model shrinkage effect, and thus result in overestimation of the number of communities as shown in~\cite{IDEPM}. To counteract this, we build a novel model using the hierarchical beta-gamma prior to prevent overfitting. In particular, the unique construction of the Dirichlet Markov chain enables us to adopt the recently advanced SG-MCMC algorithms~\cite{SGRLD,Chen,Ding,MaYiAn,ChenCY} for scalable and parallelizable inference in the proposed model.
The remainder of the paper is structured as follows. We first review relevant backgrounds. Then, we present the novel dynamic edge partition model, and describe its Gibbs sampler and SG-MCMC algorithm. We demonstrate the accuracy and efficiency of our method on several real-world datasets. Finally, we conclude the paper and discuss the future research.

\section{Preliminaries}
Throughout this paper, bold-faced lower case and upper case letters are used to denote vectors and matrices, respectively. We use ``$\boldsymbol{\cdot}$'' as an index summation shorthand, e.g., $x_{\boldsymbol{\cdot} j} = \sum_i x_{ij}$. We also use $\mathrm{Dir}()$, $\mathrm{Mult}()$, $\mathrm{DirMult}()$, $\mathrm{Gam}()$, $\mathrm{Pois}()$, $\mathrm{NB}()$, $\mathrm{Log}()$ and $\mathcal{N}()$ to denote the Dirichlet, multinomial, Dirichlet-Multinomial, gamma, Poisson, negative-binomial, logarithmic and normal distributions, respectively.

\noindent\textbf{The Poisson-Logarithmic Bivariate distribution}~\cite{AugCon, NBP}\\
If a negative-binomial distributed variable $x\sim\mathrm{NB}(r,p)$ is represented under its compound Poisson distribution as $x\sim\sum_{t=1}^l \mathrm{Log}(p), l\sim\mathrm{Pois}(-r\log(1-p))$, then the conditional distribution of $l$ has the probability mass function (PMF) as
\begin{align}
f(l\mid x, r) = \frac{\Gamma(r)}{\Gamma(x+r)}|s(x,r)|r^l,\quad l = 0,1,\ldots, x\notag,
\end{align}
where $s(x,r)$ denotes Strling number of the first kind. This random variable is referred as the Chinese Restaurant Table (CRT) distribution~\cite{csp}.

Hence, under the compound Poisson representation, the joint distribution of $x\sim\mathrm{NB}(r,p)$  and $l\sim\mathrm{CRT}(x,r)$ can be equivalently re-expressed as
\begin{align}
x\sim\mathrm{SumLog}(p),\quad l\sim\mathrm{Pois}(-r\log(1-p)),\notag
\end{align}
where $x\sim\mathrm{SumLog}(p)$ denotes the sum-logarithmic distribution generated as $x\sim\sum_{t=1}^{l}u_t,\ \ u_t\sim\mathrm{Log}(p)$.
\\
\noindent\textbf{Negative-Binomial augmentation}~\cite{zhou2018}\\
Let $r_{\boldsymbol{\cdot}} = \sum_{\scriptscriptstyle i} r_i$ and $\bm{x} = \{x_1,\ldots, x_N\}$ be random variables distributed as $\{x_1,\ldots, x_N\}\sim\mathrm{DirMult}(x_{\boldsymbol{\cdot}}, r_1, \ldots, r_N)$ where $x_{\boldsymbol{\cdot}}\sim\mathrm{NB}(r_{\boldsymbol{\cdot}}, p)$, we can equivalently sample $\{x_i\}_{i=1}^{N}$ as
$x_i\sim\mathrm{NB}(r_{i},p),\ \text{for}\ i = 1,\ldots, N$.

Hence, given Dirichlet-Multinomial distributed random variables $\{x_1,\ldots, x_N\}\sim \mathrm{DirMult}(x_{\boldsymbol{\cdot}}, r_1, \ldots, r_N)$ where $r_i\sim\mathrm{Gam}(a_i,b_i)$, our aim is to sample  the parameter $\{r_i\}_{i=1}^{N}$. To this end, we can introduce an auxiliary variable $p$ as $p\sim\mathrm{Beta}(x_{\boldsymbol{\cdot}}, r_{\boldsymbol{\cdot}})$, and then we have $x_i\sim\mathrm{NB}(r_{\boldsymbol{\cdot}},p)$. We further augment $x_i$ with a CRT distributed random variable $l_i\sim\mathrm{CRT}(x_i, r_i)$, and then according to the Poisson-Logarithmic bivariate distribution, we obtain
\begin{align}
x_i\sim\mathrm{SumLog}(p),\quad l_i\sim\mathrm{Pois}[-r_i\log(1-p)]. \notag
\end{align}
Then, the conditional distribution of $r_i$ can be easily obtained using the gamma-Poisson conjugacy.

\noindent\textbf{Stochastic Gradient MCMC}\\
Stochastic gradient MCMC (SG-MCMC) is an approximate MCMC algorithm that subsamples the data, and uses the stochastic gradients to update the parameters of interest at each step. Given a dataset $X = \{\bm{x}_i\}_{i=1}^N$, we have a generative model $p(X\mid\bm{\theta})$ where $\bm{\theta} \in\mathbb{R}^d$ is drawn from the prior $p(\bm{\theta})$. Our aim is to compute the posterior of $\bm{\theta}$, i.e., $p(\bm{\theta}\mid X) \propto \exp(-H(\theta))$ with potential function $H(\bm{\theta})\equiv -\sum_{\bm{x}_i\in X} \log p(\bm{x}_i\mid\bm{\theta})-\log p(\bm{\theta})$.
It has been shown~\cite{MaYiAn} that $p^s(\bm{\theta})\propto\exp(-H(\bm{\theta}))$ is a stationary distribution of the dynamics of a stochastic differential equation of the form as

\begin{align}
\mathrm{d}\bm{\theta} & = f(\bm{\theta})\mathrm{d}t + \sqrt{2\mathbf{D}(\bm{\theta})}\mathrm{d}\mathbf{W}(t), \notag
\end{align}
if  $f(\bm{\theta})$ is restricted to the following form as
\begin{align}
f(\bm{\theta}) & = [\mathbf{D}(\bm{\theta}) + \mathbf{Q}(\bm{\theta})]\nabla {H}(\bm{\theta}) + \Gamma(\bm{\theta}),\notag\\
\Gamma_i(\bm{\theta}) & = \sum_{j=1}^{d}\frac{\partial}{\partial \bm{\theta}_j}\left( \mathbf{D}_{ij}(\bm{\theta}) + \mathbf{Q}_{ij}(\bm{\theta})\right),\label{comp_term}
\end{align}
where $f(\bm{\theta})$ is the deterministic drift, $\mathbf{W}(t)$ is $d$--dimensional Wiener process, $\mathbf{D}(\bm{\theta})$ is a positive semi-definite diffusion matrix, and $\mathbf{Q}(\bm{\theta})$ is skew-symmetric curl matrix. This leads to the following update rule used in SG-MCMC algorithms as
\begin{align}\label{SGLD}
\bm{\theta}_{t+1}  \leftarrow \bm{\theta}_t - \epsilon_t[(\mathbf{D}(\bm{\theta}_t) & + \mathbf{Q}(\bm{\theta}_t)) \nabla \tilde{H}(\bm{\theta}_t) + \Gamma(\bm{\theta}_t) ] \\ 
& +\ \mathcal{N}(\mathbf{0}, \epsilon_t(2\mathbf{D}(\bm{\theta}_t) - \epsilon_t \hat{\mathbf{B}}_t)), \notag \\   
\tilde{H}(\bm{\theta}) = -\log p(\bm{\theta}) & - \rho \sum_{\bm{x}_i\in \tilde{X}} \log p(\bm{x}_i\mid\bm{\theta}), \notag
\end{align}
where $\{\epsilon_t\}$ is a sequence of step sizes, $\tilde{X}$ is the mini-batch subsampled from the full data $X$, $\rho\equiv {|X|}/{|\tilde{X}|}$, and $\hat{\mathbf{B}}_t$ is the estimate of stochastic gradient noise variance.

As shown in~\cite{MaYiAn}, setting $\mathbf{D}(\bm{\theta}) = \mathbf{G}(\bm{\theta})^{\scriptscriptstyle -1}$ where $\mathbf{G}(\bm{\theta})$ is the Fisher information matrix, and $\mathbf{Q}(\bm{\theta}) = 0$ in Eq.(\ref{SGLD}), we obtain the update rule of the stochastic gradient Riemannian Langevin dynamics (SGRLD) as
\begin{align}\label{SGRLD}
\bm{\theta}_{t+1}  \leftarrow \bm{\theta}_t & - \epsilon_t[(\mathbf{G}(\bm{\theta}_t)^{\scriptscriptstyle -1} \nabla \tilde{H}(\bm{\theta}_t) + \Gamma(\bm{\theta}_t) ] \\ 
& +\ \mathcal{N}\left(\mathbf{0}, 2\epsilon_t \mathbf{G}(\bm{\theta}_t)^{\scriptscriptstyle -1}\right). \notag
\end{align}

\section{The Model}

Let $\{G^{\scriptscriptstyle(t)}\}_{t=1}^{T}$ be a sequence of networks or graphs, where $G^{\scriptscriptstyle(t)}\in\{0,1\}^{N\times N}$ is the network snapshot observed at time $t$ with $N$ being the number of vertices. 
An edge is present between vertices $i$ and $j$, i.e., $G_{ij}^{\scriptscriptstyle(t)} = 1$ if they are connected at time $t$. Otherwise, $G_{ij}^{\scriptscriptstyle(t)} = 0$.
We ignore self edges $G_{ii}^{\scriptscriptstyle(t)}$. Generally, the considered temporal network can be decomposed into a set of $K$ communities, where $K$ is generally unknown \emph{a priori}. In order to extract an overlapping community structure from the given network, we represent each vertex $i$ at time $t$ by a $K$--dimensional positive memberships vector $\{ {\phi}_{ik}^{\scriptscriptstyle(t)} \}_{k=1}^{K}$, and thus each of the $K$ memberships can be considered as how \emph{actively} it is involved in the corresponding community at that time.  
In temporal networks, the observed edges among vertices change over time because the association relationships of these vertices to the underlying communities are evolving~\cite{DRIFT, LFP, MGMM}.
Hence, learning an expressive and interpretable vertex representations is the key to understanding the true dynamics of the underlying relations between vertices.
Unlike most existing methods~\cite{DRIFT, LFP, MGMM} utilizing a factorial hidden Markov model to capture the evolution of \emph{binary} vertices' memberships, we use a Dirichlet Markov chain construction to allow vertex-community memberships to vary smoothly over time.
More specifically, for each active community $k$, we draw ${\bm{\phi}}_{k}^{\scriptscriptstyle(t)}$ from a Dirichlet distribution, i.e., $\{{\phi}_{ik}^{\scriptscriptstyle(t)}\}_{i=1}^N\sim \mathrm{Dir}(\eta N{\phi}_{1k}^{\scriptscriptstyle(t-1)}, \ldots, \eta N{\phi}_{Nk}^{\scriptscriptstyle(t-1)}), \ \text{for}\ t\in\{2,\dots,T\}$, where ${\phi}_{ik}^{\scriptscriptstyle(t)}$ corresponds to the membership of vertex $i$ to community $k$ at time $t$. In particular, we draw $\bm{\phi}_k^{(1)} \sim \mathrm{Dir}(\eta)$. 
The intuition behind this construction is that each community can be thought of as a distribution over the $N$ vertices (akin to a topic model).
In temporal networks, these communities are evolving over time because the memberships of their affiliated vertices are varying.
Moreover, for each community $k$, we draw an associated weight $\lambda_k\sim\mathrm{Gam}(g_k, p_k/(1-p_k))$, where $p_k\sim\mathrm{Beta}{(c_0\alpha, c_0(1-\alpha))}$, to modulate the interactions probability between any two vertices affiliated to that community. 
Note that the hierarchical beta-gamma prior for $\lambda_k$ allows inferring the appropriate number of latent communities by shrinking the redundant community weights to zeros~\cite{NBP}. 
Finally, an edge between each pair of vertices is generated using the Bernoulli-Poisson link function~\cite{BLVM, EPM} as
\begin{align}
G_{ij}^{\scriptscriptstyle(t)}&\sim \mathbf{1}(m_{ij}^{\scriptscriptstyle(t)} \geq 1),\  m_{ij}^{\scriptscriptstyle(t)}\sim \mathrm{Pois}\Big(\sum_{k=1}^{K}{\phi}_{ik}^{\scriptscriptstyle(t)}\lambda_k{\phi}_{jk}^{\scriptscriptstyle(t)}\Big),\notag
\end{align}
where ${\phi}_{ik}^{\scriptscriptstyle(t)}\lambda_k{\phi}_{jk}^{\scriptscriptstyle(t)}$ measures how \emph{strongly} vertices $i$ and $j$ are connected at time $t$ because they are both affiliated to community $k$. Hence, naturally, the probability that a pair of vertices are connected at time $t$, will be higher if the two vertices share more common communities at that time.
Note that sampling of $\{m_{ij}^{\scriptscriptstyle(t)}\}_{i,j,t}$ only needs to be performed using rejection sampler~\cite{EPM} on nonzero entries in a given network as
\begin{align}\label{ZTP}
(m_{ij}^{\scriptscriptstyle(t)}\mid-) &\sim \begin{cases} \delta(0), & \text{if}\ G_{ij}^{\scriptscriptstyle(t)}=0 \\ \mathrm{ZTP}\Big(\sum_{k=1}^{K}{\phi}_{ik}^{\scriptscriptstyle(t)}\lambda_k{\phi}_{jk}^{\scriptscriptstyle(t)}\Big), & \text{otherwise}\end{cases}
\end{align}
where $\delta(0)$ is a point measure concentrated at $0$, ZTP is the zero-truncated Poisson distribution with support only on the positive integers, and ``--'' represents all other variables
Hence, inference in this model scales linearly with the number of nonzero edges in the given network data.

The full generative construction of the proposed model is as follows:
\begin{align}
\lambda_k&\sim\mathrm{Gam}(g_k, p_k/(1-p_k)), \notag\\
p_k&\sim\mathrm{Beta}{(c_0\alpha, c_0(1-\alpha))},\notag\\
\{{\phi}_{ik}^{\scriptscriptstyle(t)}\}_{i=1}^N  &\sim \mathrm{Dir}\left(\eta N\{ {\phi}_{ik}^{\scriptscriptstyle(t-1)}\}_{i=1}^N\right), \ \text{for}\ t\in\{2,\dots,T\}\notag\\
\{{\phi}_{ik}^{\scriptscriptstyle(1)}\}_{i=1}^N  &\sim \mathrm{Dir}(\eta\mathbf{1}_N),\quad \eta\sim\mathrm{Gam}(a_0,1/b_0),\notag\\
G_{ij}^{\scriptscriptstyle(t)}&\sim \mathbf{1}(m_{ij}^{\scriptscriptstyle(t)} \geq 1),
 m_{ij}^{\scriptscriptstyle(t)}\sim \mathrm{Pois}\Big(\sum_{k=1}^{K}{\phi}_{ik}^{\scriptscriptstyle(t)}\lambda_k{\phi}_{jk}^{\scriptscriptstyle(t)}\Big).\notag
\end{align}
\begin{figure} 
  \centering
   \includegraphics[width=10cm,height=7.5cm, keepaspectratio]{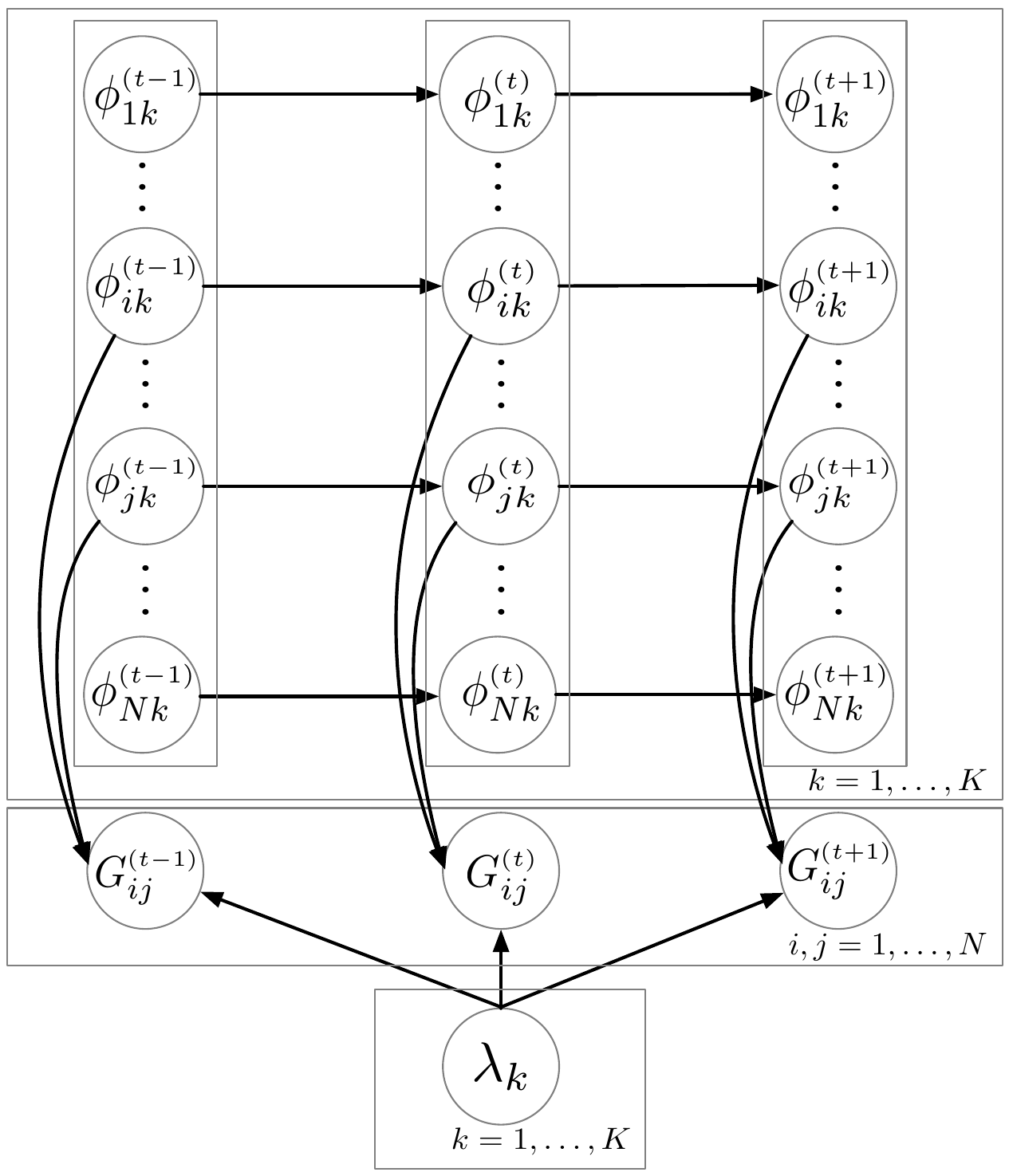}
  \caption{The plate diagram of the proposed model. Hyperpriors are omitted for brevity.}
  \label{gm}
\end{figure}
The plate diagram of the proposed model is shown in Fig.~\ref{gm}. 


%
%
\section{Inference}
Despite the proposed model not being natively conjugate for exact inference, we leverage the Negative-Binomial augmentation technique to derive a simple-to-implement Gibbs sampler with closed-form update equations. For large temporal network data, we develop a stochastic gradient MCMC algorithm using both the expanded-mean and reduced-mean re-parameterization tricks.
\subsection{Batch Gibbs Sampler}
We now proceed to describe our batch Gibbs sampler for the proposed model.
The parameters that need to be inferred are $\{m_{ij}^{\scriptscriptstyle(t)}\}_{i,j,t}, \{m_{ijk}^{\scriptscriptstyle(t)}\}_{i,j,k,t},\{\phi_{ik}^{\scriptscriptstyle(t)}\}_{i,k,t},\{\lambda_k\}_k,\{p_k\}_k\ \text{and}\ \eta$.
We present the inference update equations for each of the parameters below.
\\
\noindent\textbf{Sampling of $\{m_{ijk}^{\scriptscriptstyle(t)}\}_{i,j,t}$:} 
Using the Poisson-multinomial equivalence~\cite{NBP}, the latent counts $m_{ijk}^{\scriptscriptstyle(t)}$ are sampled from a multinomial distribution as
\begin{align}\label{eq:lcount}
\Big(\{m_{ijk}^{\scriptscriptstyle(t)}\}_{k=1}^K \mid - \Big) & \sim \mathrm{Mult}\left( m_{ij}^{\scriptscriptstyle(t)}, \frac{\{ {\phi}_{ik}^{\scriptscriptstyle(t)}\lambda_k{\phi}_{jk}^{\scriptscriptstyle(t)} \}_{k=1}^K}{\sum_{k=1}^{K}{\phi}_{ik}^{\scriptscriptstyle(t)}\lambda_k{\phi}_{jk}^{\scriptscriptstyle(t)}} \right).
\end{align}
\\
\noindent\textbf{Sampling of $\{\phi_{ik}^{\scriptscriptstyle(t)}\}_{i,k,t}$:} 
According to the additive property of the Poisson distribution, we have the aggregated counts\\ $m_{i\boldsymbol{\cdot}k}^{\scriptscriptstyle(t)}\equiv\sum_{j\neq i}m_{ijk}^{\scriptscriptstyle(t)}\ \text{and}\ m_{\boldsymbol{\cdot\cdot}k}\equiv\frac{1}{2}\sum_{i,j,t}m_{ijk}^{\scriptscriptstyle(t)}$ that can be expressed as
\begin{align}
m_{i\boldsymbol{\cdot}k}^{\scriptscriptstyle(t)}&\sim\mathrm{Pois}(\lambda_k\phi_{ik}^{\scriptscriptstyle(t)}), \\
m_{\boldsymbol{\cdot\cdot}k} &\sim\mathrm{Pois}(\lambda_kT). \label{m_dotdot}
\end{align}
Via the Poisson-multinomial equivalence, we can equivalently sample $\{m_{i\boldsymbol{\cdot}k}^{\scriptscriptstyle(t)}\}_{i=1}^N$ as
\begin{align} \label{eq:lcount2}
\{m_{i\boldsymbol{\cdot}k}^{\scriptscriptstyle(t)}\}_{i=1}^N &\sim\mathrm{Mult}\left(m_{\boldsymbol{\cdot\cdot}k}^{\scriptscriptstyle(t)}, \{\phi_{ik}^{\scriptscriptstyle(t)}\}_{i=1}^N\right).
\end{align}
For $t=T$, 
we sample the Dirichlet distributed vector $\{{\phi}_{ik}^{\scriptscriptstyle(t)}\}_{i=1}^N$ using the Dirichlet-multinomial conjugacy as
\begin{align}\label{phi_T}
(\{{\phi}_{ik}^{\scriptscriptstyle(t)}\}_{i=1}^N \mid-)  &\sim \mathrm{Dir}\left(\eta N\{{\phi}_{ik}^{\scriptscriptstyle(t-1)} + m_{i\boldsymbol{\cdot}k}^{\scriptscriptstyle(t-1)}\}_{i=1}^N\right).
\end{align}
For $2\leq t\leq (T-1)$, as we already have the multinomial likelihood and Dirichlet prior as
\begin{align}
\{m_{i\boldsymbol{\cdot}k}^{\scriptscriptstyle(t+1)}\}_{i=1}^N &\sim\mathrm{Mult}\left(m_{\boldsymbol{\cdot\cdot}k}^{\scriptscriptstyle(t+1)}, \{\phi_{ik}^{\scriptscriptstyle(t+1)}\}_{i=1}^N\right),\\
\{{\phi}_{ik}^{\scriptscriptstyle(t+1)}\}_{i=1}^N  &\sim \mathrm{Dir}\left(\eta N\{{\phi}_{ik}^{\scriptscriptstyle(t)} \}_{i=1}^N\right).
\end{align}
Marginalizing out $\{{\phi}_{ik}^{\scriptscriptstyle(t+1)}\}_{i=1}^N$ using the Dirichlet-multinomial conjugacy leads to
\begin{align}
\{m_{i\boldsymbol{\cdot}k}^{\scriptscriptstyle(t+1)}\}_{i=1}^N &\sim \mathrm{DirMult}\left(m_{\boldsymbol{\cdot\cdot}k}^{\scriptscriptstyle(t+1)}, \eta N\{{\phi}_{ik}^{\scriptscriptstyle(t)} \}_{i=1}^N\right).
\end{align}
Using the Negative-Binomial augmentation strategy, we introduce an auxiliary variable 
\begin{align}\label{zeta}
\zeta_{k}^{\scriptscriptstyle(t+1)}\sim\mathrm{Beta}(m_{\boldsymbol{\cdot\cdot}k}^{\scriptscriptstyle(t+1)}, \eta N),
\end{align}
and then the latent counts $\{m_{i\boldsymbol{\cdot}k}^{\scriptscriptstyle(t+1)}\}_{i=1}^N$ can be equivalently sampled as
\begin{align}
m_{i\boldsymbol{\cdot}k}^{\scriptscriptstyle(t+1)} &\sim \mathrm{NB}\left(\eta N {\phi}_{ik}^{\scriptscriptstyle(t)}, \zeta_{k}^{\scriptscriptstyle(t+1)}\right).\notag
\end{align}
We further augment $m_{i\boldsymbol{\cdot}k}^{\scriptscriptstyle(t+1)}$ with an auxiliary CRT distributed variable as
\begin{align}\label{xiCRT}
\xi_{ik}^{\scriptscriptstyle(t+1)}\sim\mathrm{CRT}(m_{i\boldsymbol{\cdot}k}^{\scriptscriptstyle(t+1)}, \eta N {\phi}_{ik}^{\scriptscriptstyle(t)}).
\end{align}

According to the Poisson Logarithmic bivariate distribution, we can equivalently draw $m_{i\boldsymbol{\cdot}k}^{\scriptscriptstyle(t+1)}$ and $\xi_{ik}^{\scriptscriptstyle(t+1)}$ as
\begin{align}
m_{i\boldsymbol{\cdot}k}^{\scriptscriptstyle(t+1)} &\sim \mathrm{SumLog}\left( \xi_{ik}^{\scriptscriptstyle(t+1)}, \zeta_{k}^{\scriptscriptstyle(t+1)} \right),\notag\\
\xi_{ik}^{\scriptscriptstyle(t+1)} &\sim \mathrm{Pois}\left[ -\eta N {\phi}_{ik}^{\scriptscriptstyle(t)}\log\Big( 1- \zeta_{k}^{\scriptscriptstyle(t+1)} \Big)\right].\notag
\end{align}
Via the Poisson-multinomial equivalence, we can sample $\{\xi_{ik}^{\scriptscriptstyle(t+1)}\}_{i=1}^{N}$ from a multinomial distribution as
\begin{align}\label{xi}
\{\xi_{ik}^{\scriptscriptstyle(t+1)}\}_{i=1}^N &\sim \mathrm{Mult}\left(\xi_{\boldsymbol{\cdot}k}^{\scriptscriptstyle(t+1)}, \{{\phi}_{ik}^{\scriptscriptstyle(t)} \}_{i=1}^N\right).
\end{align}
Combining the prior placed over $\{{\phi}_{ik}^{\scriptscriptstyle(t)} \}_{i=1}^N$ and the multinomial likelihood function in Eqs.(\ref{eq:lcount2};\ref{xi}), we sample $\{{\phi}_{ik}^{\scriptscriptstyle(t)} \}_{i=1}^N$ using the Dirichlet-multinomial conjugacy as
\begin{align}\label{phi_t}
(\{{\phi}_{ik}^{\scriptscriptstyle(t)}\}_{i=1}^N \mid-)  &\sim \mathrm{Dir}\left(\{\eta N{\phi}_{ik}^{\scriptscriptstyle(t-1)} + \xi_{ik}^{\scriptscriptstyle(t+1)} + m_{i\boldsymbol{\cdot}k}^{\scriptscriptstyle(t)}\}_{i=1}^N\right),
\end{align}
where $\xi_{ik}^{\scriptscriptstyle(t+1)}$ can be considered as the information passed back from time $t+1$ to $t$.
\\
\noindent\textbf{Sampling of $\eta$:}
As we already have the Poisson likelihood $\xi_{ik}^{\scriptscriptstyle(t)} \sim \mathrm{Pois}(-\eta N {\phi}_{ik}^{\scriptscriptstyle(t-1)}\log( 1- \zeta_{k}^{\scriptscriptstyle(t)}))$ and the gamma prior $\eta\sim\mathrm{Gam}(a_0,1/b_0)$, we sample $\eta$ using the gamma-Poisson conjugacy as
\begin{align}\label{eta}
(\eta\mid-)\sim\mathrm{Gam}\left(a_0 + \sum_{i,k,t}\xi_{ik}^{\scriptscriptstyle(t)}, \frac{1}{b_0 - N\sum_{k,t}[\log(1-\zeta_{k}^{\scriptscriptstyle(t)})]}\right).
\end{align}
\\
\noindent\textbf{Sampling of $\lambda_k$:}
Similaly, using the gamma-Poisson conjugacy, we obtain the conditional distribution of $\lambda_k$ as
\begin{align}\label{lambda_k}
(\lambda_k\mid-)\sim\mathrm{Gam}\left(g_k + m_{\boldsymbol{\cdot\cdot}k}, \frac{p_k}{1 + (T-1)p_k}\right).
\end{align}
\\
\noindent\textbf{Sampling of $p_k$:}
Marginalizing out $\lambda_k$ from the likelihood in Eq.(\ref{m_dotdot}) and the prior $\lambda_k\sim\mathrm{Gam}(g_k, p_k/(1-p_k))$, we obtain
$m_{\boldsymbol{\cdot\cdot}k}/T \sim \mathrm{NB}(g_k, p_k).$
Using the beta-negative-binomial conjugacy, we sample $p_k$ as
\begin{align}\label{p_k}
(p_k\mid-)\sim\mathrm{Beta}\left(c_0\alpha + m_{\boldsymbol{\cdot\cdot}k}/T , c_0(1-\alpha) + g_k\right).
\end{align}
The full inference procedure is presented  in Algorithm 1.
\begin{algorithm}[t]
\caption{Batch Gibbs Sampler for the proposed Dirichlet Dynamic Edge Partition Model}\label{alg:rejectionsampling}
\begin{algorithmic}[1]
\REQUIRE temporal graphs $\{G^{\scriptscriptstyle(t)}\}_{t}$, maximum iterations $\mathcal{J}$ 
\ENSURE posterior mean $\{\bm{\phi}_k^{\scriptscriptstyle(t)}\}_{k,t}, \{\lambda_k\}_k, \{p_k\}_k, \eta$
\FOR{$l$ = 1:$\mathcal{J}$}
	\STATE  Sample $\{m_{ij}^{\scriptscriptstyle(t)}\}_{i,j,t}$ and $\{m_{ijk}^{\scriptscriptstyle(t)}\}_{i,j,k,t}$ (Eqs.~\ref{ZTP};~\ref{eq:lcount})
	\STATE  Update $\{m_{\boldsymbol{\cdot\cdot}k}^{\scriptscriptstyle(t)}\}_{k,t}, \{m_{\boldsymbol{\cdot\cdot}k}^{\scriptscriptstyle(t)}\}_{k,t}$, and $\{m_{\boldsymbol{\cdot\cdot}k}\}_{k}$
	 \FOR{t = T,\ldots, 1}
	 \STATE Sample $\{\bm{\xi}_k^{\scriptscriptstyle(t)}\}_k\ \text{and}\ \{\zeta_k^{\scriptscriptstyle(t)}\}_k$ (Eqs.~\ref{xiCRT};~\ref{zeta})
	 \ENDFOR
	\FOR{t = 1,\ldots, T}
	\STATE Sample $\{\bm{\phi}_k^{\scriptscriptstyle(t)}\}_k$ (Eqs.~\ref{phi_T};~\ref{phi_t})
	\ENDFOR
	\STATE Sample $\{\lambda_k\}_k,\{p_k\}_k$ and $\eta$ (Eqs.~\ref{lambda_k};~\ref{p_k};~\ref{eta})
\ENDFOR
\end{algorithmic}
\end{algorithm}
\subsection{Scalable Inference via Stochastic Gradient MCMC}
While the proposed Gibbs sampler scales linearly with the number of nonzero entries in the given temporal network data, Gibbs sampler tends to be slow to mix and converge in practice. In order to mitigate this limitation, we resort to SG-MCMC algorithms for scalable inference in the proposed model.
Our SG-MCMC algorithm for the proposed model is mainly based on the stochastic gradient Riemannian Langevin dynamics for the probability simplex because of the unique construction of the Dirichlet Markov chain here. Naively applying SG-MCMC to perform inference for the probability simplex may result in invalid values being proposed. Thus, various strategies have been investigated to parameterize the probability simplex~\cite{SGRLD}.

First, we consider the expanded-mean that was shown to achieve overall best performance~\cite{SGRLD}.
In the proposed model, $\bm{\phi}_k^{\scriptscriptstyle(t)}$ is an $N$--dimensional probability simplex, and our goal is to update $\bm{\phi}_k^{\scriptscriptstyle(t)}$ as the global parameter on a mini-batch data at each step.
Using the expanded-mean trick, we parameterize $\bm{\phi}_k^{\scriptscriptstyle(t)}$ as $\{{\phi}_{1k}^{\scriptscriptstyle(t)},\ldots,{\phi}_{Nk}^{\scriptscriptstyle(t)}\} = \{\hat{\phi}_{1k}^{\scriptscriptstyle(t)},\ldots,\hat{\phi}_{Nk}^{\scriptscriptstyle(t)}\} / \hat{\phi}_{\boldsymbol{\cdot}k}^{\scriptscriptstyle(t)}$ where $\hat{\phi}_{ik}^{\scriptscriptstyle(t)}\sim\mathrm{Gam}(\eta N\phi_{ik}^{\scriptscriptstyle(t-1)}, 1)$ and $\hat{\phi}_{\boldsymbol{\cdot}k}^{\scriptscriptstyle(t)} \equiv \sum_i\hat{\phi}_{ik}^{\scriptscriptstyle(t)}$. Then, 
$\{\hat{\phi}_{1k}^{\scriptscriptstyle(t)},\ldots,\hat{\phi}_{Nk}^{\scriptscriptstyle(t)}\} / \hat{\phi}_{\boldsymbol{\cdot}k}^{\scriptscriptstyle(t)}$ 
will follow $\mathrm{Dir}\left(\eta N\{ {\phi}_{ik}^{\scriptscriptstyle(t-1)}\}_{i=1}^N\right)$ distribution.

%
%
Given the log-posterior of $\hat{\bm{\phi}}_k^{\scriptscriptstyle(t)}$ on the full data $G$ as
\begin{align}
%
\log p(\{ \hat{\phi}_{ik}^{\scriptscriptstyle(t)}\}_{i=1}^N\mid -) \propto \sum_{i=1}^N \Big[ (\tilde{m}_{ik}^{\scriptscriptstyle(t)} + \eta N {\phi}_{ik}^{\scriptscriptstyle(t-1)} - 1) \log(\hat{\phi}_{ik}^{\scriptscriptstyle(t)}) \notag\\
+ \tilde{m}_{ik}^{\scriptscriptstyle(t)}  \log(\hat{\phi}_{\boldsymbol{\cdot}k}) - \hat{\phi}_{ik}^{\scriptscriptstyle(t)} \Big]\notag
\end{align}
where $\tilde{m}_{ik}^{\scriptscriptstyle(t)}\equiv\xi_{ik}^{\scriptscriptstyle(t+1)} + m_{i\boldsymbol{\cdot}k}^{\scriptscriptstyle(t)}$, we take the gradient of the log-posterior with respect to $\hat{\bm{\phi}}_k^{\scriptscriptstyle(t)}$ on 
a mini-batch data $\tilde{G}$, and then obtain
\begin{align}\label{G01}
&\nabla_{\hat{\bm{\phi}}_k^{\scriptscriptstyle(t)}}[-\tilde{H}(\hat{\bm{\phi}}_k^{\scriptscriptstyle(t)})] 
\notag\\
& 
=  \Big\{\frac{\rho \tilde{{m}}_{ik}^{\scriptscriptstyle(t)}  + \eta N \phi_{ik}^{\scriptscriptstyle(t-1)} - {1}}{\hat{\phi}_{ik}^{\scriptscriptstyle(t)}} 
- \frac{\rho \tilde{m}_{\boldsymbol{\cdot}k}^{\scriptscriptstyle(t)} }{\hat{\phi}_{\boldsymbol{\cdot}k}^{\scriptscriptstyle(t)}} - 1 \Big\}_{i=1}^{N},
\end{align}
where $\rho\equiv|G|/|\tilde{G}|$, and $\tilde{m}_{\boldsymbol{\cdot}k}^{\scriptscriptstyle(t)}\equiv \xi_{\boldsymbol{\cdot}k}^{\scriptscriptstyle(t+1)} + m_{\boldsymbol{\cdot\cdot}k}^{\scriptscriptstyle(t)}$.

Given the gamma-Poisson construction used in expanded mean $\tilde{m}_{ik}^{\scriptscriptstyle(t)}\sim\mathrm{Pois}(\hat{\phi}_{ik}^{\scriptscriptstyle(t)})$, the Fisher information matrix is calculated as
\begin{align}\label{FIM01}
\mathrm{\mathbf{G}}\Big( \hat{\bm{\phi}}_k^{\scriptscriptstyle(t)} \Big)
& = \mathsf{E}\left\{ -\frac{\partial^2}{\partial {{\hat{{\phi}}_k^{(t)^2}}} } \log\left[ \prod_i \mathrm{Pois}\left(\tilde{m}_{ik}^{\scriptscriptstyle(t)}\ ; \hat{{\phi}}_{ik}^{\scriptscriptstyle(t)}\right) \right] \right\} \notag\\
 & = \mathrm{diag}\left({1}/{\hat{\bm{\phi}}_k^{\scriptscriptstyle(t)}}\right).  
\end{align}
Using Eq.(\ref{comp_term}), we obtain 

\begin{align}\label{comp_term_01}
\Gamma_i(\hat{\bm{\phi}}_k^{\scriptscriptstyle(t)}) &  = \sum_j \frac{\partial}{\partial \hat{{\phi}}_{kj}^{\scriptscriptstyle(t)}} \left[  \mathrm{\mathbf{G}}\Big( {\hat{\bm{\phi}}}_k^{\scriptscriptstyle(t)} \Big)^{-1} \right]_{ij} = {1}.
\end{align}

Plugging Eqs.(\ref{G01};\ref{FIM01};\ref{comp_term_01}) into Eq.(\ref{SGRLD}) yields the SGRLD update rule as\footnote{In this paper, $l$ is used to denote stepsize because $t$ is used to denote time point in temporal network data.}

\begin{align}
\Big( \hat{\phi}_{ik}^{\scriptscriptstyle(t)} \Big)^{*} = &\ \Big| \hat{\phi}_{ik}^{\scriptscriptstyle(t)} + \epsilon_l \Big[ \left(\rho \tilde{{m}}_{ik}^{\scriptscriptstyle(t)} + \eta N {\phi}_{ik}^{\scriptscriptstyle(t-1)} \right) \label{EM} \\
& - \Big(\rho \tilde{m}_{\boldsymbol{\cdot}k}^{\scriptscriptstyle(t)} + \hat{\phi}_{\boldsymbol{\cdot}k}^{\scriptscriptstyle(t)} \Big)  {\phi}_{ik}^{\scriptscriptstyle(t)}\Big]  + \mathcal{N}(0, 2\epsilon_l \hat{\phi}_{ik}^{\scriptscriptstyle(t)}) \Big|,\notag 
\\
\{{\phi}_{1k}^{\scriptscriptstyle(t)},\ldots,{\phi}_{Nk}^{\scriptscriptstyle(t)}\} = &\ \{\hat{\phi}_{1k}^{\scriptscriptstyle(t)},\ldots,\hat{\phi}_{Nk}^{\scriptscriptstyle(t)}\} / \hat{\phi}_{\boldsymbol{\cdot}k}^{\scriptscriptstyle(t)}, \notag
\end{align}
where the positiveness of $\{\hat{\phi}_{ik}^{\scriptscriptstyle(t)}\}_{i=1}^N$ is ensured by the absolute value operation $|\cdot|$. For $t=1$, the update equation is the same except that $\eta N {\phi}_{ik}^{\scriptscriptstyle(t-1)}$ is replaced by $\eta$.

Let $\bm{\psi}_{k}^{\scriptscriptstyle(t)}$ be a nonnegative vector constrained with ${\psi}_{\boldsymbol{\cdot}k}^{\scriptscriptstyle(t)}\equiv \sum_{i=1}^{N-1}{\psi}_{ik}^{\scriptscriptstyle(t)}\leq1$.
As shown in~\cite{SGRLD}, $\bm{\phi}_k^{\scriptscriptstyle(t)}$  can be alternatively parameterized via the reduced-mean trick as 
$\{{\phi}_{1k}^{\scriptscriptstyle(t)},\ldots,{\phi}_{Nk}^{\scriptscriptstyle(t)}\} =\{{\psi}_{1k}^{\scriptscriptstyle(t)},\ldots,{\psi}_{(N-1)k}^{\scriptscriptstyle(t)}, 1 - {\psi}_{\boldsymbol{\cdot}k}^{\scriptscriptstyle(t)}\}$.
Although being considered as a flawed solution because of its unstable gradients, it has been shown that this stability issue can be mitigated after preconditioning the noisy gradients~\cite{precondition}. Here, in the proposed model, we utilize the inverse of Fisher information matrix to precondition the noisy gradients, and derive an efficient update rule using the recently advanced fast sampling algorithm~\cite{cong2017}.


Given the log-posterior of $\bm{\psi}_k^{\scriptscriptstyle(t)}$ on the full data $G$ as
\begin{align}
\log p(\{{\psi}_{ik}^{\scriptscriptstyle(t)}\}_{i=1}^{N-1} \mid - ) \propto \sum_{i=1}^{N-1} (\eta N \phi_{ik}^{\scriptscriptstyle(t-1)} + \tilde{{m}}_{ik}^{\scriptscriptstyle(t)} - 1) \log({\psi}_{ik}^{\scriptscriptstyle(t)}) \notag\\
 + (\eta N \phi_{Nk}^{\scriptscriptstyle(t-1)} + \tilde{{m}}_{Nk}^{\scriptscriptstyle(t)} - 1) \log(1 - {\psi}_{\boldsymbol{\cdot} k}^{\scriptscriptstyle(t)}) \notag
\end{align}
we take the gradient of the log-posterior with respect to $\bm{\psi}_k \in \mathbb{R}_{\geq 0}^{N-1}$ on a mini-batch data scaled by $\rho \equiv {|G|}/{|\tilde{G}|}$, and then we have
\begin{align}
&\nabla_{\bm{\psi}_k^{\scriptscriptstyle(t)}}[-\tilde{H}(\bm{\psi}_k^{\scriptscriptstyle(t)})] \label{G02}\\
& =  \Big\{\frac{\rho \tilde{{m}}_{ik}^{\scriptscriptstyle(t)}  + \eta N \phi_{ik}^{\scriptscriptstyle(t-1)} - {1}}{{\psi}_{ik}^{\scriptscriptstyle(t)}} 
- \frac{\rho \tilde{m}_{Nk}^{\scriptscriptstyle(t)} + \eta N \phi_{Nk}^{\scriptscriptstyle(t-1)} - {1}}{1 - {\psi}_{\boldsymbol{\cdot}k}^{\scriptscriptstyle(t)}} \Big\}_{i=1}^{\scriptscriptstyle N-1}.\notag
\end{align}

Note that the gradient in Eq.(\ref{G02}) becomes unstable if some of the components of $\bm{\psi}_k^{\scriptscriptstyle(t)}$ approach zeros. Nevertheless, this issue can be mitigated after preconditioning the noisy gradient with the inverse of Fisher information matrix.

\begin{algorithm}[t]
\caption{Stochastic Gradient MCMC for the proposed Dirichlet Dynamic Edge Partition Model}
\begin{algorithmic}[1]
\REQUIRE temporal graphs $\{G^{\scriptscriptstyle(t)}\}_{t}$, maximum iterations $\mathcal{J}$ 
\ENSURE posterior mean $\{\bm{\phi}_k^{\scriptscriptstyle(t)}\}_{k,t}, \{\lambda_k\}_k, \{p_k\}_k, \eta$
\FOR{$l$ = 1:$\mathcal{J}$}
	\STATE  Gibbs sampling on the $l$-th mini-batch for $\{m_{ij}^{\scriptscriptstyle(t)}\}_{i,j,t}$, $\{m_{ijk}^{\scriptscriptstyle(t)}\}_{i,j,k,t}$, $\{\lambda_k\}_k,\{p_k\}_k$ and $\eta$;
	\STATE Update $\{m_{\boldsymbol{\cdot\cdot}k}^{\scriptscriptstyle(t)}\}_{k,t}, \{m_{\boldsymbol{\cdot\cdot}k}^{\scriptscriptstyle(t)}\}_{k,t}$, and $\{m_{\boldsymbol{\cdot\cdot}k}\}_{k}$ 
	\STATE /* Update global parameters */
	\FOR{t = 1,\ldots, T}
	\STATE Update $\{\bm{\phi}_k^{\scriptscriptstyle(t)}\}_k$ (Eqs.~\ref{EM}; ~\ref{RM})
	\ENDFOR
\ENDFOR
\end{algorithmic}
\end{algorithm}

Given the multinomial likelihood as
\begin{align}
\{\tilde{m}_{ik}^{\scriptscriptstyle(t)}\}_{i=1}^N &\sim \mathrm{Mult}\left(\tilde{m}_{\boldsymbol{\cdot}k}^{\scriptscriptstyle(t)}, \{{\phi}_{ik}^{\scriptscriptstyle(t)} \}_{i=1}^N\right),
\end{align}
we calculate the Fisher information matrix of $\bm{\psi}_k^{\scriptscriptstyle(t)}$ as
\begin{align}\label{FIM02}
& \mathrm{\mathbf{G}}\Big( \bm{\psi}_k^{\scriptscriptstyle(t)} \Big) \notag\\
& = \mathsf{E}\left\{ -\frac{\partial^2}{\partial {\bm{\psi}_k^{\scriptscriptstyle(t)}}^2} \log\left[ \mathrm{Mult}\left(\{\tilde{m}_{ik}^{\scriptscriptstyle(t)}\}_{i=1}^N\ ;\ \tilde{m}_{\boldsymbol{\cdot}k}^{\scriptscriptstyle(t)}, \{{\phi}_{ik}^{\scriptscriptstyle(t)} \}_{i=1}^N\right) \right] \right\} \notag\\
 & = M_k^{\scriptscriptstyle(t)}\left\{ \mathrm{diag}\left(\frac{1}{\bm{\psi}_k^{\scriptscriptstyle(t)}}\right) + \frac{\mathbf{11^T}}{1 - {\psi}_{\boldsymbol{\cdot}k}^{\scriptscriptstyle(t)}} \right\},
\end{align}
where $M_k^{\scriptscriptstyle(t)} \equiv \mathsf{E}{[ \tilde{m}_{\boldsymbol{\cdot}k}^{\scriptscriptstyle(t)} ]}$. 
Using Eq.(\ref{comp_term}), we have 
\begin{align}\label{comp_term_02}
\Gamma_i(\bm{\psi}_{k}^{\scriptscriptstyle(t)}) &  = \sum_j \frac{\partial}{\partial {\psi}_{kj}^{\scriptscriptstyle(t)}} \left[  \mathrm{\mathbf{G}}\Big( \bm{\psi}_k^{\scriptscriptstyle(t)} \Big)^{-1} \right]_{ij} = (1 - N{\psi}_{ik}^{\scriptscriptstyle(t)})/M_k^{\scriptscriptstyle(t)}.
\end{align}

Substituting Eqs.(\ref{G02};\ref{FIM02};\ref{comp_term_02}) in Eq.(\ref{SGRLD}), we obtain the following SGRLD update rule as
\begin{align}\label{update_rule_02}
&\Big( \bm{\psi}_{k}^{\scriptscriptstyle(t)} \Big)^{*} \notag\\
& = \left\{ \bm{\psi}_{k}^{\scriptscriptstyle(t)} + \frac{\epsilon_l}{\scriptstyle M_k^{\scriptscriptstyle(t)}} \Big[ \left(\rho \tilde{\mathbf{m}}_{k}^{\scriptscriptstyle(t)} + \eta N {\bm{\tilde{\phi}}}_{k}^{\scriptscriptstyle(t-1)} \right) - \Big(\tilde{m}_{\boldsymbol{\cdot}k}^{\scriptscriptstyle(t)} + \eta N \Big)  \bm{\psi}_{k}^{\scriptscriptstyle(t)}\Big]  \right. \notag\\
& \left. +\ \mathcal{N}\left(0, \frac{2\epsilon_l}{\scriptstyle M_k^{\scriptscriptstyle(t)}} \left[ \mathrm{diag}(\bm{\psi}_{k}^{\scriptscriptstyle(t)}) - \bm{\psi}_{k}^{\scriptscriptstyle(t)}{\bm{\psi}_{k}^{\scriptscriptstyle(t)}}^{\mathrm{T}} \right]\right) \right\}_{\angle},
\end{align}
where ${\bm{\tilde{\phi}}}_{k}^{\scriptscriptstyle(t)} \equiv [{\phi}_{1k}^{\scriptscriptstyle(t)}, \ldots, {\phi}_{(N-1)k}^{\scriptscriptstyle(t)}]$, and $\{\cdot\}_{\angle}$ denotes the constraint that ${\psi}_{ik}^{\scriptscriptstyle(t)} \geq 0$ and $\sum_{i=1}^{N-1}{\psi}_{ik}^{\scriptscriptstyle(t)} \leq 1$.

It is computational expensive to simulate the multivariate normal distribution in Eq.(\ref{update_rule_02}) using Cholesky decomposition. Therefore, we resort to a recently advanced fast sampling algorithm~\cite{cong2017}. Instead of updating $\bm{\psi}_{k}^{\scriptscriptstyle(t)}$, we can equivalently update $\bm{\phi}_{k}^{\scriptscriptstyle(t)}$ that is drawn from a related multivariate normal distribution with a diagonal covariance matrix as
\begin{align}
&\Big( \bm{\phi}_{k}^{\scriptscriptstyle(t)} \Big)^{*} \notag\\
& = \left\{ \bm{\phi}_{k}^{\scriptscriptstyle(t)} + \frac{\epsilon_l}{\scriptstyle M_k^{\scriptscriptstyle(t)}} \Big[ \left(\rho \tilde{\mathbf{m}}_{k}^{\scriptscriptstyle(t)} + \eta N {\bm{\tilde{\phi}}}_{k}^{\scriptscriptstyle(t-1)} \right) - \Big(\tilde{m}_{\boldsymbol{\cdot}k}^{\scriptscriptstyle(t)} + \eta N \Big)  \bm{\phi}_{k}^{\scriptscriptstyle(t)}\Big]  \right. \notag\\
& \left. +\ \mathcal{N}\left(0, \frac{2\epsilon_l}{\scriptstyle M_k^{\scriptscriptstyle(t)}} \left[ \mathrm{diag}(\bm{\phi}_{k}^{\scriptscriptstyle(t)})\right]\right) \right\}_{\angle}. \label{RM}
\end{align}
For $t=1$, we replace $\eta N {\bm{\tilde{\phi}}}_{k}^{\scriptscriptstyle(t-1)}$ by $\eta$ in the update rule.
Our SG-MCMC algorithm iteratively updates the parameters $\{\bm{\phi}_{k}^{\scriptscriptstyle(t)}\}_{k,t}$ and samples the remaining ones as in the proposed Gibbs sampler. The main procedure is summarized in Algorithm 2.

\section{Related Work}

Dynamic network models (DNMs) are mainly categorized into matrix factorization based methods~\cite{MTF},  the exponential random graph models~\cite{ERGM} and probabilistic methods~\cite{DRIFT, EMMB, LFP, MGMM}. The probabilistic DNMs allow us to predict missing links using the generative process, and facilitate model selection, and thus received significant attention. 
The proposed model falls in the class of the probabilistic DNMs, and we hence confine ourselves to this category. 
The representative probabilistic DNMs include the dynamic {mixed membership models}~\cite{Fu09, xing2010, EMMB,DSBM}, and the dynamic latent feature relational models (LFRM)~\cite{DRIFT,LFP,MGMM}. 
These methods show competitive performance in terms of link predictions although inference in these models require MCMC sampling that scales poorly to large datasets. The edge partition model (EPM)~\cite{EPM} and its extensions~\cite{TopicKG, Piyush17, DPGM} leverage the Bernoulli-Poisson link to generate edges, and inference in EPMs only needs to be performed on nonzero edges. Hence, the EPMs are readily scalable to moderate-sized networks with thousands of vertices. 
It is challenging to apply these probabilistic DNMs to handle the increasing amount of graph-structured data constantly created in dynamic environments. In recent years, scalable inference algorithms, such as stochastic variational inference~\cite{SVI} and stochastic gradient MCMC~\cite{SGRLD,ScalingDTM,SGMCMC_MMSBM, DLDA} have been developed for topic models, mixed membership stochastic block models etc. In this work, we developed a SG-MCMC algorithm for the new proposed dynamic edge partition model, and compared it against state-of-the-art dynamic stochastic block model~\cite{DSBM} and the gamma process dynamic network model~\cite{DPGM}.
We note that the hierarchical beta-gamma prior has been utilized in the beta-negative-binomial process (BNBP) topic model~\cite{BNBP}. The Dirichlet Markov chain is hinted by Theorem 1 of~\cite{zhou2018} and first realized in dynamic topic model~\cite{DualMarkovChain}.
To the best of our knowledge, this is the first attempt to apply these techniques in dynamic network modelling.

\section{Experiments}
 \begin{table*}[!htbp]
\small
\centering
\begin{tabular}{|l|c|c|c|c|c|}\hline
Method & Hypertext & Blog & Facebook Like & Facebook Message & NIPS Co-authorship \\
\hline
DSBM & ${0.703} $ & ${0.830}$ & 0.848 & ${0.814}$ & ${0.899}$  \\\hline
GaP-DNM & ${0.766} $ & ${0.864}$ & 0.887 &  ${0.888}$ & ${0.887}$  \\\hline
$\text{D}^2$EPM-Gibbs & $\bm{0.812} $ & $\mathbf{0.927}$ & $\mathbf{0.912}$ &  $\bm{0.929}$ & ${0.895}$  \\\hline
$\text{D}^2$EPM-EM-SGRLD & ${0.808} $ & ${0.882}$ & 0.871 &  ${0.926}$ & ${0.902}$  \\\hline
$\text{D}^2$EPM-RM-SGRLD & ${0.809} $ & ${0.875}$ & 0.868 & ${0.927}$ & $\bm{0.916}$  \\\hline
\end{tabular}
\vspace{-0.75em}
\caption{\label{Prediction} {Link prediction on temporal network data. We report the averaged area under the ROC curve (AUROC) over five different training/test partitions, and highlight the best scores in bold.}}
\end{table*}
We now present the experimental results on several real-world data sets to evaluate the accuracy and efficiency of the proposed model. 
The proposed model is referred to as {$\text{D}^2$EPM} (the \textbf{D}irichlet \textbf{D}ynamic \textbf{E}dge \textbf{P}artition \textbf{M}odel) with Gibbs sampler, Expanded-Mean SGRLD and Reduced-Mean SGRLD, as {$\text{D}^2$EPM}-Gibbs, {$\text{D}^2$EPM-EM-SGRLD, {$\text{D}^2$EPM-RM-SGRLD, respectively.
We compare our model with two baselines: (1) the dynamic stochastic block model (DSBM)~\cite{DSBM}.
(2) the gamma process dynamic network model (GaP-DNM) that captures the evolution of vertices' memberships using a gamma Markov chain construction~\cite{DPGM}. 
We also compare the proposed SG-MCMC algorithms in terms of link prediction accuracy vs wall-clock run time. 
We chose the following datasets in our experiments:
\noindent\textbf{(1) Hypertext:} 
This dataset~\cite{ffdc_data} contains the interactions between 113 participants at the 2009 Hypertext conference.
We generated a dynamic network
assuming each hour as a snapshot, and creating an
edge between each pair of participants at time $t$ if they have at
least one contact recorded during that snapshot.
\noindent\textbf{(2) Blog:} 
The blog dataset~\cite{YangTB}, contains $148,681$ edges among 407 blogs during 15 months.
\noindent\textbf{(3) Facebook Like\footnote{\url{https://tinyurl.com/ycdezko6}.}:} 
This dataset contains 33,720 broadcast messages among 899 students
over 7 months from a Facebook like forum. We generated a dynamic network aggregating the data into monthly snapshots, and creating
an edge between each pair of vertices if the presence of messages between them is recorded during that snapshot.
\noindent\textbf{(4) Facebook Message\footnote{\url{https://snap.stanford.edu/data/CollegeMsg.html}.}:} 
This dataset contains 59,835 private messages among 1,899 college students
over 7 months. We generated a dynamic network aggregating the data into monthly snapshots, and creating
an edge between each pair of vertices if the presence of messages between them is recorded during that snapshot.
\noindent\textbf{(5) NIPS Co-authorship\footnote{\url{http://www.cs.huji.ac.il/~papushado/nips_collab_data.html}.}:} 
This dataset contains 4,798 publications by 5,722 authors in the NIPS conference over 10 years. We generated
a dynamic network aggregating the data into yearly snapshots and creating an edge between two authors in a snapshot if they appear on the same publication in that year.
We note that only medium-scale network data sets are chosen in our experiments because it is infeasible to run MCMC sampling based methods on large network data. 

First, we compare the accuracy of all the models in terms of link prediction. We train all the methods using $80\%$ of randomly chosen entries (either links or non-links) in the given network data, and use the remaining $20\%$ as the held-out data to test the trained model. Each experiment is conduced five times with different training/test partitions, and the averaged Area Under the Receiver Operating Characteristi curve (AUROC) for all the data sets is reported in the final results.
The proposed method is implemented in MATLAB.
Unless specified otherwise, we initialize the GaP-DNM and the proposed {$\text{D}^2$EPM} with $K = 50$ because both two models can automatically determine the number of communities. We run both GaP-DNM and  {$\text{D}^2$EPM}-Gibbs for 3000 iterations with 2000 burn-in and 1000 collection iterations. We set the hyperparameters as $g_k = 0.1, a_0= b_0 = 0.01, c_0 = 1$ and $\alpha = 1/K$. A sensitivity analysis revealed that we obtain similar results when instead setting $g_k = 0.01$ or $1$.
The SG-MCMC algorithms were also run for the same number of iterations, with mini-batch size equal to one-fourth of the number of nonzero edges in the training data. We use the stepsize $\epsilon_l = (a(1+l/b))^{-c}$ and the optimal parameters $a, b, c$ as in~\cite{SGRLD,MaYiAn}.
All the experiments are conducted on a standard computer with 24 GB RAM.
%
Table 1 shows the experimental results on the link prediction task. Overall, the {$\text{D}^2$EPM} (both Gibbs sampling and SG-MCMC algorithms) outperform the other baselines. Specifically, the sampling based methods (GaP-DNM and {$\text{D}^2$EPM}-Gibbs) achieve better accuracy than the DSBM based on the extended Kalman filter on the relatively small datasets although the former two methods require a sufficiently large number of iterations to converge. 
For the medium-sized NIPS dataset, the SG-MCMC algorithms perform better than the Gibbs sampling, suggesting the batch Gibbs sampler mixes poorly.
In Figure~\ref{runtime}, we compare the AUROC vs wall-clock run time for {$\text{D}^2$EPM}-Gibbs, {$\text{D}^2$EPM-EM-SGRLD, {$\text{D}^2$EPM-RM-SGRLD on Facebook message and NIPS datasets. For Facebook message dataset, we found that batch Gibbs sampler converges very fast, and the SG-MCMC algorithms converges to the same level of accuracy in comparable time. For the larger network (NIPS dataset), the SG-MCMC algorithms converge much faster than the Gibbs sampler. 
\vspace{-0.75em}
\begin{figure} 
  \centering
   \begin{tabular}{ c c } 
   \hspace{-1.25em}
    \begin{minipage}{.2105\textwidth}
      \includegraphics[width=4.5cm,height=4.5cm, keepaspectratio, angle=0]{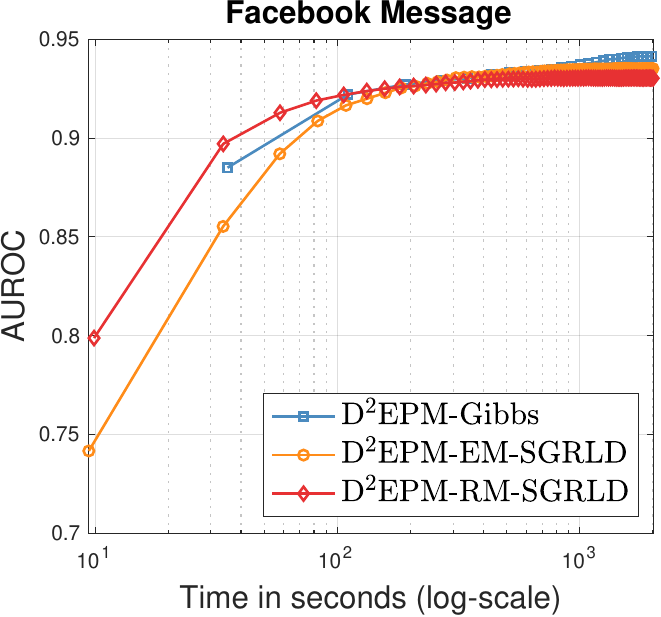}
    \end{minipage}
    &
     \hspace{5em}
      \begin{minipage}{.2105\textwidth}
      \includegraphics[width=4.5cm,height=4.5cm, keepaspectratio, angle=0]{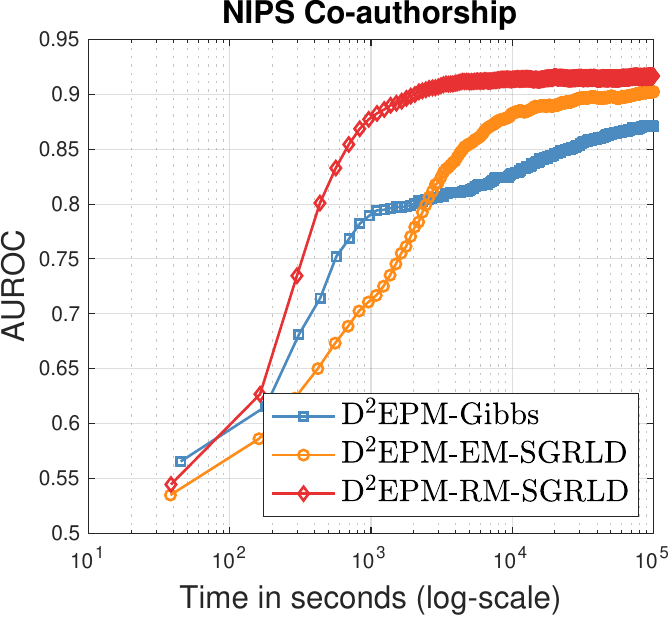}
    \end{minipage}
  \end{tabular}
  \vspace{-0.75em}
  \caption{Running time comparison of {$\text{D}^2$EPM}-Gibbs, $\text{D}^2$EPM-EM-SGRLD, $\text{D}^2$EPM-RM-SGRLD on Facebook message (left) and NIPS Co-authorship datasets (right).}
  \label{runtime}
\end{figure}

\section{Conclusion}
We presented a novel dynamic edge partition model for temporal relational learning by capturing the evolution of vertices' memberships over time using a Dirichlet Markov chain construction. The appropriate number of latent communities is automatically inferred via the hierarchical beta-gamma prior. In particular, the new framework admits a simple-to-implement Gibbs sampling scheme using the Negative-Binomial augmentation technique, and also enables us to develop a scalable inference algorithm based on the SG-MCMC framework. We demonstrate the accuracy and efficiency of the novel methods on several real-world datasets.
The proposed framework allows us to incorporate available vertex-specific side information via the P\'olya-Gamma augmentation technique~\cite{PG}, and also to infer a tree-structured latent communities hierarchy using the gamma belief-net~\cite{AGBN}. As more aspects of social interactions are collected, there is an increasing demand for privacy-preserving learning methods. It would be interesting to investigate the privatizing Bayesian inference~\cite{PBI} for the proposed method. In the future research, we plan to study stochastic variational inference scheme\cite{8594893} and Wasserstein barycentre based sampling~\cite{JMLR:v19:17-084,SDM-23} for the Poisson-gamma dynamic network models.

\bibliographystyle{plain}
\bibliography{main}

\begin{thebibliography}{10}

\bibitem{DualMarkovChain}
Ayan Acharya et~al.
\newblock A dual {M}arkov chain topic model for dynamic environments.
\newblock In {\em KDD}, pages 1099--1108, 2018.

\bibitem{MMSBM}
Edoardo~M. Airoldi et~al.
\newblock Mixed membership stochastic blockmodels.
\newblock {\em J. Mach. Learn. Res.}, 9:1981--2014, June 2008.

\bibitem{ScalingDTM}
Arnab Bhadury et~al.
\newblock Scaling up dynamic topic models.
\newblock In {\em WWW}, pages 381--390, 2016.

\bibitem{KnowBase}
Antoine Bordes et~al.
\newblock Learning structured embeddings of knowledge bases.
\newblock In {\em AAAI}, pages 301--306, 2011.

\bibitem{Charlin}
Laurent Charlin et~al.
\newblock Dynamic {P}oisson factorization.
\newblock In {\em RecSys}, pages 155--162, 2015.

\bibitem{ChenCY}
Changyou Chen et~al.
\newblock Stochastic gradient mcmc with {S}tale gradients.
\newblock In {\em NIPS}, pages 2937--2945, 2016.

\bibitem{Chen}
Tianqi Chen et~al.
\newblock {Stochastic Gradient {H}amiltonian {M}onte {C}arlo}.
\newblock In {\em ICML}, pages 1683--1691, 2014.

\bibitem{DLDA}
Yulai Cong et~al.
\newblock Deep latent {D}irichlet allocation with topic-layer-adaptive
  stochastic gradient {R}iemannian {MCMC}.
\newblock In {\em ICML}, pages 864--873, 2017.

\bibitem{cong2017}
Yulai Cong et~al.
\newblock Fast simulation of hyperplane-truncated multivariate normal
  distributions.
\newblock {\em Bayesian Analysis}, 12(4):1017--1037, 12 2017.

\bibitem{Ding}
Nan Ding et~al.
\newblock {B}ayesian sampling using stochastic gradient thermostats.
\newblock In {\em NIPS}, pages 3203--3211, 2014.

\bibitem{Cao18}
Trong Dinh~Thac Do and Longbing Cao.
\newblock Coupled {P}oisson factorization integrated with user/item metadata
  for modeling popular and sparse ratings in scalable recommendation.
\newblock In {\em AAAI}, pages 2918--2925, 2018.

\bibitem{MTF}
Daniel~M. Dunlavy et~al.
\newblock Temporal link prediction using matrix and tensor factorizations.
\newblock {\em ACM Trans. Knowl. Discov. Data}, 5(2):10:1--10:27, 2011.

\bibitem{BLVM}
David~B. Dunson et~al.
\newblock {Bayesian} latent variable models for mixed discrete outcomes.
\newblock {\em Biostatistics}, 6(1):11--25, 2005.

\bibitem{DRIFT}
James~R. Foulds et~al.
\newblock A dynamic relational infinite feature model for longitudinal
  networks.
\newblock In {\em AISTATS}, pages 287--295, 2011.

\bibitem{Fu09}
Wenjie Fu et~al.
\newblock Dynamic mixed membership blockmodel for evolving networks.
\newblock In {\em ICML}, pages 329--336, 2009.

\bibitem{pnas_a}
M.~Girvan and M.~E.~J. Newman.
\newblock Community structure in social and biological networks.
\newblock {\em Proceedings of the National Academy of Sciences},
  99(12):7821--7826, 2002.

\bibitem{Gopalan}
Prem Gopalan et~al.
\newblock Scalable recommendation with hierarchical poisson factorization.
\newblock In {\em UAI}, pages 326--335, 2015.

\bibitem{LFP}
Creighton Heaukulani et~al.
\newblock Dynamic probabilistic models for latent feature propagation in social
  networks.
\newblock In {\em ICML}, pages 275--283, 2013.

\bibitem{EMMB}
Qirong Ho et~al.
\newblock Evolving cluster mixed-membership blockmodel for time-evolving
  networks.
\newblock In {\em AISTATS}, pages 342--350, 2011.

\bibitem{SVI}
Matthew~D. Hoffman et~al.
\newblock Stochastic variational inference.
\newblock {\em JMLR}, 14:1303--1347, 2013.

\bibitem{TopicKG}
Changwei Hu et~al.
\newblock Topic-based embeddings for learning from large knowledge graphs.
\newblock In {\em AISTATS}, pages 1133--1141, 2016.

\bibitem{DGMR}
Changwei Hu et~al.
\newblock Deep generative models for relational data with side information.
\newblock In {\em ICML}, pages 1578--1586, 2017.

\bibitem{MGMM}
Myunghwan Kim et~al.
\newblock Nonparametric multi-group membership model for dynamic networks.
\newblock In {\em NIPS}, pages 1385--1393, 2013.

\bibitem{precondition}
Chunyuan Li et~al.
\newblock Preconditioned stochastic gradient {L}angevin dynamics for deep
  neural networks.
\newblock In {\em AAAI}, pages 1788--1794, 2016.

\bibitem{SGMCMC_MMSBM}
Wenzhe Li et~al.
\newblock Scalable mcmc for mixed membership stochastic blockmodels.
\newblock In {\em AISTATS}, pages 723--731, Cadiz, Spain, 2016.

\bibitem{MaYiAn}
Yi-An Ma et~al.
\newblock A complete recipe for stochastic gradient {MCMC}.
\newblock In {\em NIPS}, pages 2917--2925, 2015.

\bibitem{ffdc_data}
Rossana Mastrandrea et~al.
\newblock {C}ontact patterns in a high school: {A} comparison between data
  collected using wearable sensors.
\newblock {\em PLoS ONE}, 10(9):1--26, 2015.

\bibitem{LFRM}
Kurt Miller et~al.
\newblock Nonparametric latent feature models for link prediction.
\newblock In {\em NIPS}, pages 1276--1284, 2009.

\bibitem{KnowGraph}
Maximilian Nickel et~al.
\newblock A review of relational machine learning for knowledge graphs.
\newblock {\em Proc. of the {IEEE}}, 104(1):11--33, 2016.

\bibitem{IDEPM}
Iku Ohama et~al.
\newblock On the model shrinkage effect of gamma process edge partition models.
\newblock In {\em NIPS}, pages 397--405, 2017.

\bibitem{SGRLD}
Sam Patterson et~al.
\newblock Stochastic gradient {R}iemannian {L}angevin dynamics on the
  probability simplex.
\newblock In {\em NIPS}, pages 3102--3110, 2013.

\bibitem{csp}
J.~Pitman.
\newblock {\em Combinatorial stochastic processes}.
\newblock Springer-Verlag, Berlin, 2006.
\newblock Lectures on Probability Theory.

\bibitem{PG}
Nicholas~G. Polson et~al.
\newblock {B}ayesian inference for logistic models using {P}{\'o}lya--{G}amma
  latent variables.
\newblock {\em JASA}, 108(504):1339--1349, 2013.

\bibitem{Piyush17}
Piyush Rai.
\newblock Non-negative inductive matrix completion for discrete dyadic data.
\newblock In {\em AAAI}, pages 2499--2505, San Francisco, USA, 2017.

\bibitem{ERGM}
Garry Robins et~al.
\newblock An introduction to exponential random graph (p*) models for social
  networks.
\newblock {\em Social Networks}, 29(2):173--191, May 2007.

\bibitem{BPTD}
Aaron Schein et~al.
\newblock {B}ayesian {P}oisson {T}ucker decomposition for learning the
  structure of international relations.
\newblock In {\em ICML}, pages 2810--2819, 2016.

\bibitem{PBI}
Aaron Schein et~al.
\newblock Locally private {B}ayesian inference for count models.
\newblock {\em CoRR}, abs/1803.08471, 2018.

\bibitem{JMLR:v19:17-084}
Sanvesh Srivastava, Cheng Li, and David~B. Dunson.
\newblock Scalable {B}ayes via barycenter in wasserstein space.
\newblock {\em Journal of Machine Learning Research}, 19(8):1--35, 2018.

\bibitem{NDKG}
Yi~Tay et~al.
\newblock Non-parametric estimation of multiple embeddings for link prediction
  on dynamic knowledge graphs.
\newblock In {\em AAAI}, pages 1243--1249, 2017.

\bibitem{xing2010}
Eric~P. Xing et~al.
\newblock A state-space mixed membership blockmodel for dynamic network
  tomography.
\newblock {\em Ann. Appl. Stat.}, 4(2):535--566, 06 2010.

\bibitem{DSBM}
Kevin~S. Xu et~al.
\newblock Dynamic stochastic blockmodels for time-evolving social networks.
\newblock {\em J. Sel. Topics Signal Processing}, 8(4):552--562, 2014.

\bibitem{8594893}
Sikun Yang and Heinz Koeppl.
\newblock Collapsed variational inference for nonparametric {B}ayesian group
  factor analysis.
\newblock In {\em 2018 IEEE International Conference on Data Mining (ICDM)},
  pages 687--696, 2018.

\bibitem{ICML-18}
Sikun Yang and Heinz Koeppl.
\newblock Dependent relational gamma process models for longitudinal networks.
\newblock In {\em Proceedings of the International Conference on Machine
  Learning (ICML)}, pages 5551--5560, 2018.

\bibitem{DPGM}
Sikun Yang and Heinz Koeppl.
\newblock A {P}oisson gamma probabilistic model for latent node-group
  memberships in dynamic networks.
\newblock In {\em AAAI}, pages 4366--4373, 2018.

\bibitem{UAI-20}
Sikun Yang and Heinz Koeppl.
\newblock The {H}awkes edge partition model for continuous-time event-based
  temporal networks.
\newblock In {\em Proceedings of the 36th Conference on Uncertainty in
  Artificial Intelligence (UAI)}, pages 460--469, 2020.

\bibitem{SDM-23}
Sikun Yang and Hongyuan Zha.
\newblock Estimating latent population flows from aggregated data via inversing
  multi-marginal optimal transport.
\newblock In {\em Proceedings of the 2023 SIAM International Conference on Data
  Mining (SDM)}, pages 181--189, 2023.

\bibitem{YangTB}
Tianbao Yang et~al.
\newblock Detecting communities and their evolutions in dynamic social networks
  - a {B}ayesian approach.
\newblock {\em Machine Learning}, 82(2):157--189, 2011.

\bibitem{Zhao}
He~Zhao et~al.
\newblock Leveraging node attributes for incomplete relational data.
\newblock In {\em ICML}, pages 4072--4081, 2017.

\bibitem{BNBP}
Mingyuan Zhou.
\newblock Beta-negative binomial process and exchangeable random partitions for
  mixed-membership modeling.
\newblock In {\em NIPS}, pages 3455--3463, 2014.

\bibitem{EPM}
Mingyuan Zhou.
\newblock Infinite edge partition models for overlapping community detection
  and link prediction.
\newblock In {\em AISTATS}, pages 1135--1143, 2015.

\bibitem{zhou2018}
Mingyuan Zhou.
\newblock Nonparametric {B}ayesian negative binomial factor analysis.
\newblock {\em {B}ayesian Analysis}, pages 1--29, 2018.

\bibitem{AugCon}
Mingyuan Zhou and Lawrence Carin.
\newblock Augment-and-conquer negative binomial processes.
\newblock In {\em NIPS}, pages 2555--2563, 2012.

\bibitem{NBP}
Mingyuan Zhou and Lawrence Carin.
\newblock Negative binomial process count and mixture modeling.
\newblock {\em {IEEE} Trans. PAMI}, 37(2):307--320, 2015.

\bibitem{AGBN}
Mingyuan Zhou et~al.
\newblock Augmentable gamma belief networks.
\newblock {\em Journal of Machine Learning Research}, 17:1--44, 2016.

\end{thebibliography}

\end{document}